\documentclass[english,aps,manuscript,preprintnumbers]{revtex4-1}
\usepackage[T1]{fontenc}
\usepackage[latin9]{inputenc}
\usepackage{amsmath}
\usepackage{graphicx}
\usepackage{esint}

\makeatletter

\@ifundefined{textcolor}{}
{%
 \definecolor{BLACK}{gray}{0}
 \definecolor{WHITE}{gray}{1}
 \definecolor{RED}{rgb}{1,0,0}
 \definecolor{GREEN}{rgb}{0,1,0}
 \definecolor{BLUE}{rgb}{0,0,1}
 \definecolor{CYAN}{cmyk}{1,0,0,0}
 \definecolor{MAGENTA}{cmyk}{0,1,0,0}
 \definecolor{YELLOW}{cmyk}{0,0,1,0}
}

\makeatother

\usepackage{babel}
\begin{document}

\preprint{BROWN-HET-1671}

\title{Holographic operator mapping in dS/CFT and cluster decomposition}

\author{Atreya Chatterjee}

\author{David A. Lowe}

\email[To whom correspondence should be addressed: ]{lowe@brown.edu}

\affiliation{Department of Physics, Brown University, Providence, RI 02912, USA}
\begin{abstract}
The bulk to boundary mapping for massive scalar fields is constructed,
providing a de Sitter analog of the LSZ reduction formula. The set
of boundary correlators thus obtained defines a potentially new class
of conformal field theories based on principal series representations
of the global conformal group. Conversely, we show bulk field operators
in de Sitter may be reconstructed from boundary operators. While consistent
at the level of the free field theory, the boundary CFT does not satisfy
cluster decomposition. The resulting conformal field theory does not
satisfy the basic axioms of Euclidean quantum field theory due to
Osterwalder and Schrader, so is likely not well-defined once interactions
are included.
\end{abstract}
\maketitle

\section{Introduction}

The Bekenstein-Hawking entropy \citep{beken,Hawking:1974sw} 
\[
S=\frac{A}{4G}\,,
\]
states that the entropy of any black hole is proportional to its surface
area. This is law widely applicable in various kinds of spacetime.
This law suggests that any theory in the bulk can be described in
terms of some boundary theory of spacetime in one lesser dimension. 

There has been a lot of progress in understanding this correspondence
between bulk quantum theory in anti-de Sitter spacetime and boundary
conformal field theory \citep{Aharony:1999ti}. One expects these
ideas to carry over in some form to the cases of asymptotically flat
spacetime and asymptotically de Sitter spacetime. In these cases the
situation is much less clear, and our aim in the present work is to
carefully set up the bulk/boundary correspondence in the de Sitter
case. This will allow us to draw some interesting conclusions about
the structure of the novel conformal field theories that must appear
in this case, and their ultimate consistency.

Various different formulations of dS/CFT has been proposed, following
Strominger's initial work \citep{Strominger:2001pn}. Some formulations
simply extend the AdS/CFT correspondence to the dS space via analytic
continuation, which has been successful for massless fields (and possibly
sub-Hubble mass fields) and the massless higher spin gravity theories
\citep{Anninos:2011ui}. Our goal in the present work is to investigate
the situation for generic fields with masses larger than the Hubble
scale, which are related by analytic continuation to tachyonic fields
in anti-de Sitter spacetime. New methods must be developed to treat
this case. It is worth noting that in the CFT these fields will correspond
to quasi-primary fields with complex conformal weights. Nevertheless,
these form unitary representations of the global conformal group \citep{Chernikov:1968zm,Tagirov:1972vv,Guijosa:2003ze,Guijosa:2005qi},
opening the door to possibility that an entirely new class of conformal
field theories might be defined based on these representations.

One of the key mysteries in the dS/CFT correspondence is the origin
of bulk time, since the dual CFT is a purely Euclidean theory. In
the AdS/CFT correspondence this is not an issue because the bulk time
is parallel to the boundary time and the CFT lives in a spacetime
with Lorentzian signature. As a result, it becomes more interesting
to see how unitarity and time ordering in the bulk theory emerges
from the Euclidean CFT, and we will obtain partial results in this
direction. 

The paper is organized as follows. We begin by presenting an analog
of the LSZ construction \citep{Lehmann:1954rq} for quantum fields
in de Sitter spacetime, which provides a clear definition of correlators
in the boundary CFT. This step is necessary because the representations
of the conformal group in question, the principal series, are not
commonly studied in the context of conformal field theory. The construction
is inspired by the integral geometry approach of Gelfand \citep{gelfand66},
and many of the results detailed there carry over to the present case.
For the most part, our focus will be on three-dimensional de Sitter
spacetime, though many of the ideas carry over to the higher dimensional
case.

We then consider the inverse map, reconstructing bulk field operators
in terms of the CFT data. At leading order (essentially the free level
from the viewpoint of quantum field theory in the bulk) we construct
bulk creation and annihilation field operators using operators in
the CFT. Bulk operator ordering in correlators can be accomplished
by adopting an $i\epsilon$-prescription, complexifying the radial
direction in the CFT. This is sufficient to recover the bulk Wightman
two-point correlation function, with the correct Hadamard singularity
at light-like separations. This approach may also be used to build
higher point correlators, for bulk theories with perturbative expansions,
by using the creation and annihilation operators to reproduce the
Wick expansion. However a completely general nonperturbative understanding
of the bulk operator ordering, and hence the origin of bulk time,
is elusive. 

The construction we describe allows one to define a CFT from some
set of bulk correlators in de Sitter spacetime. We may then proceed
to analyze the basic consistency of the resulting CFT, to check whether
it satisfies the basic axioms expected of a Euclidean quantum field
theory. These are known as the Osterwalder-Schrader axioms \citep{Osterwalder:1973dx,Osterwalder:1974tc}.
One of these axioms is the Euclidean version of cluster decomposition,
which requires correlators to factorize in the limit of large separations.
We find this fails in the case of the principal series, if, for example,
operators of the form $L_{1}\bar{L}_{1}\mathcal{O}_{\Delta}$ are
considered, where $L_{1}$ and $\bar{L}_{1}$ are conformal generators
that raise the weight by $1$, and $\mathcal{O}_{\Delta}$ is a quasi-primary
operator with weight $\Delta$. Note in ordinary CFTs with $\mathcal{O}_{\Delta}$
a primary field with positive conformal weight, the combination $L_{1}\mathcal{O}_{\Delta}$
would vanish. The operator $L_{1}\bar{L}_{1}\mathcal{O}_{\Delta}$
will be dual to a graviton plus a massive matter field insertion.
The failure of cluster decomposition signals that the vacuum of the
CFT is not unique, i.e. there can be many excitations in the bulk
that give rise to nontrivial operators on the boundary satisfying
$L_{0}=\bar{L}_{0}=0$. This follows from the lack of a positive energy
theorem in the bulk theory \citep{Abbott:1981ff},%
\footnote{There is a positive energy theorem for the global timelike conformal
Killing vector of de Sitter \citep{Kastor:2002fu}, but it is not
clear if this is well-defined on the conformal compactification of
de Sitter. So its relation to the dual CFT is not currently understood.%
}. We note this does not immediately imply infrared divergences in
the bulk theory. In fact, the classical stability of de Sitter spacetime
for pure gravity or massless conformal matter coupled to gravity has
been demonstrated \citep{FRIEDRICH:1986fk,Friedrich:1986kx,Friedrich:1991nn}.
Most likely, this result should be interpreted as an incompleteness
in the CFT dual to an interacting theory in de Sitter spacetime, a
point we hope to return to in future work.

A related construction of bulk operators from boundary operators in
dS/CFT has been considered in \citep{Xiao:2014uea,Sarkar:2014dma}.
There are numerous differences in the details and conclusions with
the present work.

\section{Basic setup\label{sec:Imaginary-Lobachevskian-Spaces}}

In this section we will introduce some notation, broadly following
the integral geometry approach of \citep{gelfand66} in imaginary
Lobachevskian space, also known as elliptic de Sitter spacetime \citep{Folacci:1986gr,Parikh:2002py}.
Elliptic de Sitter is simply global de Sitter modulo the antipodal
map. Our main focus will be global de Sitter. In some ways elliptic
de Sitter is simpler because there is a single connected boundary
at infinity, while in global de Sitter there are two disconnected
boundaries, one in the distant past, and one in the distant future.
We will find in global de Sitter a CFT may be defined on either boundary,
and for the sake of definiteness we choose the past boundary. Our
formulas will be explicitly written for the case of three-dimensional
de Sitter spacetime, but the results generalize immediately to higher
dimensions.

The de-Sitter space can be realized on a hyperboloid embedded in four-dimensional
Minkowski spacetime 

\begin{equation}
x_{0}^{2}-x_{1}^{2}-x_{2}^{2}-x_{3}^{2}=-R^{2}\,,\label{eq:hyperbol}
\end{equation}
where $R$ is some positive constant. The geodesic distance $r$ between
any two points 
\begin{equation}
\cosh^{2}kr=\frac{\left\langle x,y\right\rangle ^{2}}{\left\langle x,x\right\rangle \left\langle y,y\right\rangle }\,,\label{eq:distance}
\end{equation}
where 
\[
\left\langle x,y\right\rangle =x_{0}y_{0}-x_{1}y_{1}-x_{2}y_{2}-x_{3}y_{3}\,,
\]
is the inner product of two vectors and $k=\frac{1}{R}$ is another
positive constant.

The family of points satisfying $\left\langle x,x\right\rangle =-R^{2}$
with antipodal points ($x\sim-x$) identified is called imaginary
Lobachevskian space or elliptic de Sitter spacetime. Without the identification
we have ordinary global de Sitter spacetime. The distance $r$ can
be real and non-negative (if $1\leq\cosh kr\leq\infty$) or imaginary
in the interval $[0,\frac{\pi i}{2k}]$ (if $0\leq\cosh kr\leq1$).
Any point on the light cone in the embedding space will be denoted
by $\xi$, that is $[\xi,\xi]=0$. 

Now let us consider some surfaces in de Sitter with particularly simple
transformation properties under the isometry group. The equation describing
a sphere of radius $r$ with center at $a$ is given by 
\[
\left\langle x,a\right\rangle {}^{2}=c\left\langle a,a\right\rangle \left\langle x,x\right\rangle \,.
\]

Consider taking the center to the infinity while ensuring that the
sphere passes through a fixed point $b$. The surface obtained in
this way is called a horosphere. In this limit, the product $c\left\langle a,a\right\rangle $
is fixed to some constant $c_{1}$ to obtain the surface

\begin{equation}
\left\langle x,\xi\right\rangle {}^{2}=c_{1}\left\langle x,x\right\rangle \,.\label{eq:horosphere}
\end{equation}

When $c_{1}<0$ this is called a horosphere of the first kind. It
is possible to normalize $c_{1}=-1$ by normalizing $\xi$. If we
set $R=1$, so that $\left\langle x,x\right\rangle =-1$ then the
horospheres of the first kind look like

\begin{equation}
|\left\langle x,\xi\right\rangle |=1\,.\label{eq:horosphere1}
\end{equation}
Thus a horosphere of the first kind may be specified by choosing a
point $\xi$ on the positive cone, $\left\langle \xi,\xi\right\rangle =0\,,\xi_{0}>0$.

When $c_{1}=0$ one gets a horosphere of the second kind

\begin{equation}
\left\langle x,\xi\right\rangle =0\,.\label{eq:horosphere2}
\end{equation}
In this paper, our focus will be on the horospheres of the first kind,
which will correspond to principal series representations of the de
Sitter group \citep{gelfand66}. We consider further the horospheres
of the second kind, which correspond to the discrete series representations,
in future work.

\section{Boundary CFT operators\label{sec:Boundary-CFT-operators}}

It is useful to begin by reviewing the decomposition of some general
bounded, normalizable function on de Sitter into components that transform
as unitary irreducible representations of the conformal group \citep{gelfand66}.
For every $f(x)$ one constructs the integral transform 

\begin{equation}
h(\omega)=\int_{\omega}f(x)d\sigma\,,\label{eq:integral}
\end{equation}
and $d\sigma$ is an invariant measure, and the integral is over a
horosphere of first kind $\omega$. Let the equation of a horosphere
be $|\left\langle x,\xi\right\rangle |=1$. Equation~\eqref{eq:integral}
can also be written as

\begin{equation}
h(\xi)=\int f(x)\delta\left(|\left\langle x,\xi\right\rangle |-1\right)dx\,,\label{eq:transform}
\end{equation}
where $dx$ is the invariant measure on the de Sitter spacetime. This
above map is nothing but a generalization of the Fourier transform,
which takes a function defined on the horosphere to a function defined
on the lightcone labelled by $\xi$. As we will see, $\xi$ can be
used to parametrize the boundary at past infinity in de Sitter. 

Now consider functions $h(\xi)$ over the positive sheet of the light
cone where $\xi^{0}>0$. These functions may be decomposed into components
with well-defined conformal weights by Fourier transforming

\begin{equation}
F(\xi;\rho)=\int_{0}^{\infty}h(t\xi)t^{-i\rho}dt\,,\label{eq:fourier-1}
\end{equation}
where the complex conformal weight $\Delta$ is related to the real
parameter $\rho$ via $i\rho=1-\Delta$. Let us note that inserting
\eqref{eq:transform} into \eqref{eq:fourier-1} we have
\[
F(\xi;\rho)=\int_{0}^{\infty}dt\int dxf(x)\delta\left(\left|\left\langle x,t\xi\right\rangle \right|-1\right)t^{-i\rho}\,.
\]
Performing the integral over $t$ we arrive at
\begin{equation}
F(\xi;\rho)=\int dxf(x)\left|\left\langle x,\xi\right\rangle \right|^{-\Delta}\,.\label{eq:bfunction}
\end{equation}

Generalizing $f$ to some bulk correlator of some scalar field of
mass $m$, our goal will then be to view the analog of $F$ as a boundary
correlator. A key difference with the work of Gelfand is that we must
give up the condition of normalizability (in the sense that $\int|f(x)|^{2}dx$
is finite). As we will see, this the de Sitter isometry covariant
component of \eqref{eq:transform} will correspond to the residue
of a pole in $\rho^{2}-m^{2}$ reminiscent of the LSZ reduction formula
in flat spacetime \citep{Lehmann:1954rq}.

\subsection{Flat slicing}

\begin{figure}
\includegraphics{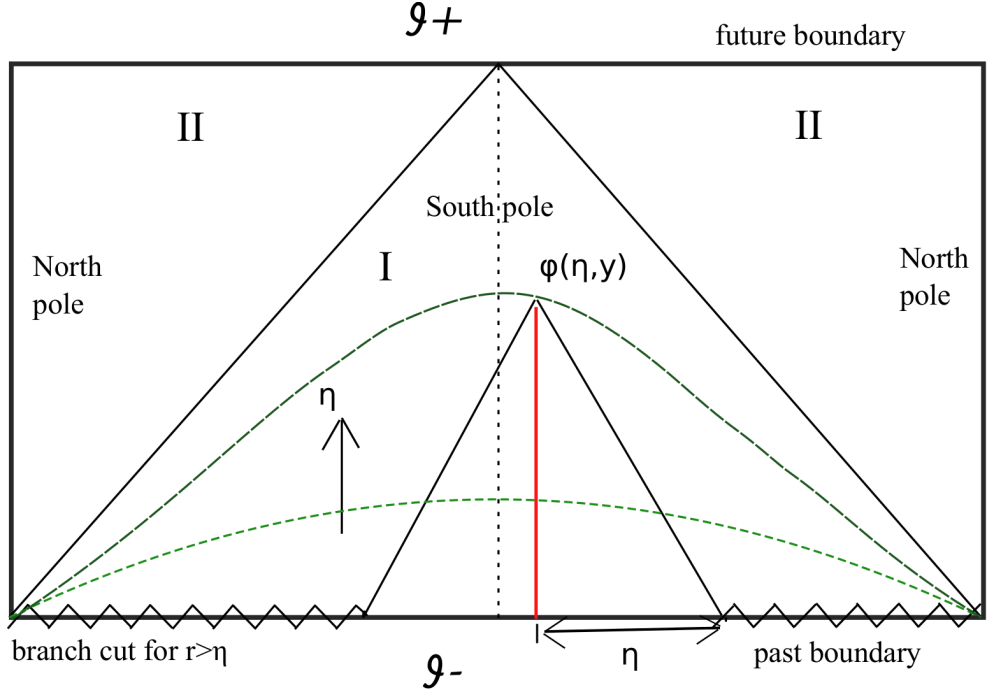}\protect\caption{\label{fig:Penrose-diagram-for}Penrose diagram for de-Sitter space.
The vertical dashed line is the South pole. The left and right edges
are North pole which are identified with each other. Horizontal dashed
lines are constant $\eta$ slices. The figure shows the calculation
of bulk field from boundary operator in the past boundary. The boundary
operator is smeared over the whole boundary. There is branch cut in
the smearing function for $r>\eta$. Continuing $r$ and $\eta$ to
the complex plane via an $i\epsilon$ prescription selects the branch
yielding a Bunch-Davies/Euclidean vacuum positive or negative frequency
mode.}

\end{figure}

Horospheres of the first kind are diffeomorphic to flat spatial slices
in de Sitter. It therefore will be convenient to express the general
coordinate invariant expression \eqref{eq:bfunction} on flat slices.
See \citep{Drechsler:1978cw} for some related work in the context
of four-dimensional de Sitter. Setting $R=1$, the 3-dimensional de
Sitter hyperboloid can be parameterized by the coordinates $(\eta,y_{1},y_{2})$
via
\begin{eqnarray*}
x^{0} & = & \frac{1}{2}(\eta-\frac{1}{\eta})-\frac{\sum y_{i}^{2}}{2\eta}\\
x^{1} & = & \frac{y_{1}}{\eta}\\
x^{2} & = & \frac{y_{2}}{\eta}\\
x^{3} & = & -\frac{1}{2}(\eta+\frac{1}{\eta})+\frac{\sum y_{i}^{2}}{2\eta}\,,
\end{eqnarray*}
yielding the de Sitter metric with a flat spatial slicing and conformal
time $\eta$
\[
ds^{2}=\frac{d\eta^{2}}{\eta^{2}}-\frac{1}{\eta^{2}}\left(dy_{1}^{2}+dy_{2}^{2}\right)\,.
\]
The volume measure is
\begin{equation}
dx=\frac{1}{\eta^{3}}d\eta dy_{1}dy_{2}\,.\label{eq:volumeform-1}
\end{equation}
A point on a light cone may be parameterized by 
\begin{equation}
\xi=k\lambda(1+z^{2},2z_{1},2z_{2},1-z^{2})\,,\label{eq:lightcone-1}
\end{equation}
where $z^{2}=z_{1}^{2}+z_{2}^{2}$. The coordinates $z_{1},z_{2}$
label a point on the boundary at past infinity in de Sitter. In these
coordinates we have
\[
\left\langle x,\xi\right\rangle =\lambda\eta\left(1-\frac{(y_{1}+z_{1})^{2}}{\eta^{2}}-\frac{(y_{2}+z_{2})^{2}}{\eta^{2}}\right)\,.
\]
We will also need the measure on the cone 
\[
d\xi=\frac{d\xi_{1}d\xi_{2}d\xi_{3}}{\xi_{0}}\,,
\]
and the measure on the boundary at infinity
\[
d\omega=d^{2}z\,.
\]

\subsection{Transform from bulk to boundary}

Our aim is to use the transform of Gelfand \citep{gelfand66} as a
guide to constructing the transform for the class of functions that
appear in correlation functions of quantum fields in de Sitter. In
particular, these functions do not satisfy the compact support condition
used in Gelfand's inversion theorem. This will lead us to build an
analog of the flat-spacetime LSZ reduction formula for de Sitter spacetime,
requiring some important differences with Gelfand's construction.

We begin by noting the mode expansion for a bulk scalar field of mass
$m$ \citep{Birrell:1982ix}
\begin{equation}
\phi(\eta,y)=c_{1}\int\frac{d^{2}k}{(2\pi)^{2}}\left(a_{k}\eta H_{i\mu}^{(2)}\left(|k|\eta\right)e^{ik.y}+a_{k}^{\dagger}\eta H_{i\mu}^{*(2)}\left(|k|\eta\right)e^{-ik.y}\right)\,,\label{eq:modeexpansion-2}
\end{equation}
where $c_{1}=\frac{\sqrt{\pi}}{2}e^{\frac{\pi\mu}{2}},$ $\mu=\sqrt{m^{2}-1}$,
and $H_{i\mu}^{(2)}(|k|\eta)$ are Hankel functions of second kind.
The operators $a_{k}$ and $a_{k}^{\dagger}$ are annihilation and
creation operators, with the $a_{k}$ annihilating the Bunch-Davies
vacuum, and
\[
[a_{k},a_{k'}^{\dagger}]=(2\pi)^{2}\delta^{(2)}(k-k')\,.
\]

To construct the boundary operator, we perform the following integral
over region I of figure \ref{fig:Penrose-diagram-for},

\begin{eqnarray}
\Phi_{\Delta}(z) & = & c_{1}\int\frac{d^{2}k}{(2\pi)^{2}}(a_{k}\eta H_{i\mu}^{(2)}(|k|\eta)e^{ik.y}+a_{k}^{\dagger}\eta H_{i\mu}^{*(2)}(|k|\eta)e^{-ik.y})\nonumber \\
 &  & \left(1-\frac{(y_{1}+z_{1})^{2}}{\eta^{2}}-\frac{(y_{2}+z_{2})^{2}}{\eta^{2}}\right)^{-\Delta}\eta^{-(3+\Delta)}d\eta d^{2}y\,.\label{eq:boundop}
\end{eqnarray}
We define the cut in the $x^{-\Delta}$ factor as
\[
x^{-\Delta}=|x|^{-\Delta}e^{-i\Delta\arg x}\,,
\]
where $\arg x\in(-\pi,\pi]$. Note this choice of phase differs from
the expression \eqref{eq:bfunction} and will be related to the choice
of the Bunch-Davies/Euclidean vacuum for the free theory. Other phase
conventions can lead to the more general $\alpha$-vacua \citep{Allen:1985ux}
which are thought to be unphysical \citep{Goldstein:2003ut}.

At the level of the bulk correlators, the operator ordering is determined
by continuing the bulk time $\eta\to\eta\pm i\epsilon$. This then
yields the distinctive signature of the Hadamard singularity of the
two-point correlator in the light-like limit, which in turn matches
the short-distance singularities of flat-spacetime \citep{Spradlin:2001pw}.
This continuation determines the branch of the cut in \eqref{eq:boundop},
and as we will see projects onto the $a_{k}$ or the $a_{k}^{\dagger}$
terms dependent on the sign. Therefore we define $P_{\Delta}$ and
$P_{\Delta}^{\dagger}$ as follows
\begin{eqnarray*}
P_{\Delta}(z) & = & \Phi_{\Delta}(z)\,,\qquad\eta\to\eta+i\epsilon\\
P_{\Delta}^{\dagger}(z) & = & \Phi_{\Delta}(z)\,,\qquad\eta\to\eta-i\epsilon
\end{eqnarray*}
with $\epsilon>0$. Performing the integrals we then get
\begin{eqnarray*}
P_{\Delta}(z) & = & d(\Delta)\frac{1}{(\Delta-1)^{2}+\mu^{2}}\int\frac{d^{2}k}{(2\pi)^{2}}a_{k}|k|^{-1+\Delta}e^{ik\cdot z}\\
P_{\Delta}^{\dagger}(z) & = & \tilde{d}(\Delta)\frac{1}{(\Delta-1)^{2}+\mu^{2}}\int\frac{d^{2}k}{(2\pi)^{2}}a_{k}^{\dagger}|k|^{-1+\Delta}e^{-ik\cdot z}\,,
\end{eqnarray*}
where
\begin{eqnarray*}
d(\Delta) & = & i2^{2-\Delta}e^{-i\pi\Delta/2}\sqrt{\pi}\Gamma(1-\Delta)\sin\left(\pi\Delta\right)\\
\tilde{d}(\Delta) & = & -i2^{2-\Delta}e^{i\pi\Delta/2}\sqrt{\pi}\Gamma\left(1-\Delta\right)\sin\left(\pi\Delta\right)
\end{eqnarray*}

We note the prefactors of the boundary operators have poles when $\Delta=1\pm i\mu$,
reminiscent of the poles arising in momentum space when one performs
the LSZ reduction in flat spacetime, which yields the S-matrix. In
the same way, we find by taking the residues of these poles, we are
able to define conformally covariant operators on the boundary
\begin{equation}
\mathcal{O}_{\Delta}(z)=d(\Delta)\frac{i}{2(\Delta-1)}\int\frac{d^{2}k}{(2\pi)^{2}}a_{k}|k|^{-1+\Delta}e^{ik\cdot z}\,,\label{eq:bop}
\end{equation}
where now $\Delta=1-i\mu$. The other pole yields the operator $\mathcal{O}_{2-\Delta}(z)$.
As we will see there is an equivalence between these two operators,
since either may be used to reconstruct the bulk annihilation mode.
A similar relation is found in the work of Gelfand. For the principal
series, the representations corresponding to $\Delta$ and $2-\Delta$
are equivalent, so the minimal spectrum of representations corresponds
to $\mu>0$. The formulas carry over straightforwardly to the operators
$\mathcal{O}_{\Delta}^{\dagger}$ and $\mathcal{O}_{2-\Delta}^{\dagger}$
.

Using this construction, we may then build the boundary two-point
correlators from the bulk Wightman function by plugging into \eqref{eq:boundop}.
The bulk Wightman function is \citep{Spradlin:2001pw}
\begin{equation}
G_{E}(x,x')=\frac{\Gamma(\Delta)\Gamma(2-\Delta)}{(4\pi)^{3/2}\Gamma\left(\frac{3}{2}\right)}\,_{2}F_{1}\left(\Delta,2-\Delta;\frac{3}{2};\frac{1+\left\langle x,x'\right\rangle }{2}\right)\,,\label{eq:bulktwo}
\end{equation}
where $x$ and $x'$ are complexified to give the correct $i\epsilon$
prescription near the light-like singularity. This may also be written
in terms of the integral over mode functions as
\[
G_{E}(x,x')=c_{1}^{2}\int\frac{d^{2}k}{(2\pi)^{2}}\eta H_{i\mu}^{(2)}\left(|k|\eta\right)\eta'H_{i\mu}^{(2)*}\left(|k|\eta'\right)e^{ik.(y-y')}\,.
\]
Performing the bulk to boundary transform on each mode function, and
taking residues yields the non-vanishing two-point correlators 
\begin{eqnarray*}
\left\langle \mathcal{O}_{\Delta}(z)\mathcal{O}_{\Delta}^{\dagger}(0)\right\rangle  & = & -\frac{\pi\sin\left(\pi\Delta\right)}{\left(\Delta-1\right)^{2}}\frac{1}{|z|^{2\Delta}}\\
\left\langle \mathcal{O}_{2-\Delta}(z)\mathcal{O}_{2-\Delta}^{\dagger}(0)\right\rangle  & = & \frac{\pi\sin\left(\pi\Delta\right)}{\left(\Delta-1\right)^{2}}\frac{1}{|z|^{2(2-\Delta)}}\,.
\end{eqnarray*}
It is helpful to recall that scalings and translations fix the form
of the correlator, but only covariance under inversions gives the
requirement that each operator in the two-point function have the
same conformal weight. Potential off-diagonal contributions vanish
as required when the integrals \eqref{eq:boundop} are performed.

The operators $\mathcal{O}_{\Delta}$, etc. are quasi-primary operators,
in the sense that they transform under $SL(2,C)$ transformations
\begin{eqnarray*}
z & \to & \frac{\alpha z+\beta}{\gamma z+\delta},\qquad\alpha\delta-\beta\gamma=1\\
\mathcal{O}_{\Delta}(z) & \to & \left|\gamma z+\delta\right|^{-2\Delta}\mathcal{O}_{\Delta}\left(\frac{\alpha z+\beta}{\gamma z+\delta}\right)\,.
\end{eqnarray*}
Note, however, that in the principal series, they are not annihilated
by the positive weight generators of $SL(2,C)$. Thus $L_{1}\mathcal{O}_{\Delta}\neq0$
and $\bar{L}_{1}\mathcal{O}_{\Delta}\neq0$ so that the operators
are not primary operators. The only representations of the conformal
group that behave as the usual CFT primary operators are the discrete
series.

The appearance of $\mathcal{O}_{\Delta}$ and $\mathcal{O}_{\Delta}^{\dagger}$
as separate operators in the CFT is somewhat unusual. The Hermitian
conjugation is not the natural one typically used in conformal field
theory, but rather refers to bulk Hermitian conjugation with respect
to the Klein-Gordon inner product. Likewise, it is with respect to
this bulk inner product, the one typically used in quantum field theory
in curved spacetime, that the representations are unitary.

Having performed this construction for a single set of de Sitter mode
functions, and the two-point function, one can try to generalize to
higher point functions. As is clear from the above discussion, the
residue of the integral transform \eqref{eq:boundop} essentially
picks off a free ingoing or outgoing mode, depending on the branch
of the integrand the $i\epsilon$ term picks. Therefore, if the bulk
quantum field theory satisfies cluster decomposition, one may apply
the transform to a multi-point correlation function to define a de
Sitter version of the S-matrix, in analogy with the LSZ reduction
formula. The resulting S-matrix should transform covariantly under
global conformal transformations. As we will see shortly, the existence
of this S-matrix will hinge on this assumption of cluster decomposition.

\section{Reconstructing the Bulk}

It is helpful to again recall the integral geometry construction of
\citep{gelfand66}. Having constructed the boundary function $h(\xi)$,
the bulk function is reconstructed by the inverse transform 

\begin{equation}
f(x)=-\frac{1}{16\pi^{2}}\int\delta''\left(\left|[x,\xi]\right|-1\right)h(\xi)d\xi\,,\label{eq:inversetransform}
\end{equation}
where the measure $d\xi$ is described in more detail in \citep{gelfand66}.
This can also be written as

\begin{equation}
f(x)=-\frac{1}{16\pi^{2}}\int_{0}^{\infty}\delta''(t-1)H(x,t)dt=-\frac{1}{16\pi^{2}}H_{t}''(x,1)\,,\label{eq:invtrans2}
\end{equation}
where 
\[
H(x,t)=\int h(\xi)\delta\left(\left|[x,\xi]\right|-t\right)d\xi\,.
\]

Now consider functions $h(\xi)$ over the positive sheet of the light
cone. These functions may be decomposed into components with well-defined
conformal weights by Fourier transforming

\begin{equation}
F(\xi;\rho)=\int_{0}^{\infty}h(t\xi)t^{-i\rho}dt\,,\label{eq:fourier}
\end{equation}
where the complex conformal weight $\Delta$ is related to the real
parameter $\rho$ via $i\rho=1-\Delta$. The inverse Fourier transform
becomes 

\begin{equation}
h(\xi)=\frac{1}{8\pi}\int_{-\infty}^{\infty}F(\xi;\rho)d\rho\,.\label{eq:fourierinv}
\end{equation}

Using equation \eqref{eq:fourier} and \eqref{eq:fourierinv} we get

\begin{equation}
f(x)=-\frac{1}{2(4\pi)^{3}}\int_{-\infty}^{\infty}\int F(\xi;\rho)\delta''\left(|[x,\xi]|-1\right)d\xi d\rho\,.
\end{equation}
This can be written in the form

\begin{equation}
f(x)=\frac{1}{4(8\pi)^{3}}\int_{-\infty}^{\infty}d\rho\,\rho(\rho+4i)\int_{\Gamma}d\omega\, F(\xi;\rho)\left|[x,\xi]\right|{}^{-i\rho-1}\,,\label{eq:composed}
\end{equation}
where $d\omega$ is a measure on the boundary at infinity, obtained
by modding out the overall scale from $d\xi$. The surface $\Gamma$
is an arbitrary surface on the light-cone that intersects each of
its generators, and $d\omega$ is defined by $d\xi=d\omega dP$ where
$P(\xi)=1$ is the equation of $\Gamma$. Thus we get a function in
the bulk by applying the inverse integral transform to functions on
the boundary transforming with well-defined conformal weights. Finally,
a symmetry of this integral relates the integral over $\rho$ from
$-\infty,0$ to the range $0,\infty$, allowing the range to be collapsed
to one copy of each irreducible principal series representation $\rho=0\cdots\infty$.

\subsection{Bulk operators}

Again we will need to generalize these methods to the distributions
encountered in quantum field theory. Our goal is to reconstruct the
bulk field, at the free level \eqref{eq:modeexpansion-2} using only
the covariant boundary operators \eqref{eq:bop}. For simplicity we
assume only a single mass field with mass $m$ is present. Generalization
to the quasi-free case, where a superposition of masses is present
is straightforward. The inverse transform of $\mathcal{O}_{\Delta}$
, in the flat-slicing, is
\[
\phi_{-}(\eta,y)=-\frac{\left(\Delta-1\right)^{2}}{2\pi^{2}}\left(\cot\pi\Delta+i\right)\int d^{2}z\left(\frac{\eta^{2}-z^{2}}{\eta}\right)^{\Delta-2}\mathcal{O}_{\Delta}(z+y)\,.
\]
The continuation of $\eta\to\eta-i\epsilon$ defines the branch of
the integrand. Likewise we define

\[
\phi_{+}(\eta,y)=\frac{\left(\Delta-1\right)^{2}}{2\pi^{2}}\left(\cot\pi\Delta-i\right)\int d^{2}z\left(\frac{\eta^{2}-z^{2}}{\eta}\right)^{\Delta-2}\mathcal{O}_{\Delta}^{\dagger}(z+y)\,,
\]
where now $\eta\to\eta+i\epsilon$. Inserting the expression \eqref{eq:bop}
and performing the integrals, one recovers \eqref{eq:modeexpansion-2}
with $\phi=\phi_{+}+\phi_{-}$.

The same method may be used to reconstruct the bulk Wightman function
in the Bunch-Davies/Euclidean vacuum
\begin{eqnarray*}
\left\langle \phi\left(\eta_{1},y_{1}\right)\phi\left(\eta_{2},y_{2}\right)\right\rangle  & = & -\frac{\left(\Delta-1\right)^{4}}{4\pi^{4}}\csc^{2}\left(\pi\Delta\right)\int d^{2}z_{1}\left(\frac{\eta_{1}^{2}-z_{1}^{2}}{\eta_{1}}\right)^{\Delta-2}\int d^{2}z_{2}\left(\frac{\eta_{2}^{2}-z_{2}^{2}}{\eta_{2}}\right)^{\Delta-2}\\
 & \times & \left\langle \mathcal{O}_{\Delta}(z_{1}+y_{1})\mathcal{O}_{\Delta}^{\dagger}(z_{2}+y_{2})\right\rangle \,,
\end{eqnarray*}
where on the right-hand-side a CFT correlator appears, while on the
left, a bulk Wightman function appears. In this formula, it is understood
that $\eta_{1}\to\eta_{1}-i\epsilon$ and $\eta_{2}\to\eta_{2}+i\epsilon$.
Likewise the boundary radial directions must be continued in the same
way, which regulates the singularity in the integrand when points
coincide. We emphasize this reproduces the full bulk Wightman function
for general points in the bulk of de Sitter \eqref{eq:bulktwo}. 

This construction allows us to build field operators at arbitrary
bulk points in de Sitter yielding important insight into how the de
Sitter time arises from the purely Euclidean CFT. Likewise, the Euclidean
CFT does not have a natural operator ordering. In the bulk, this arises
from the complexification of the radial direction in the CFT, combined
with the branch choices in the smearing functions. This allows us
to build ingoing or outgoing modes in the bulk. For a bulk theory
with some perturbative expansion, this approach is sufficient to reconstruct
the bulk correlators from the boundary correlators, by reconstructing
the Wick expansion of the bulk correlators, using the building blocks
we have presented.

\section{Euclidean axioms}

For a well-defined set of bulk correlators, we can use the prescription
of section \ref{sec:Boundary-CFT-operators} to define a conformally
covariant set of boundary correlators. These then may be viewed as
a definition of some Euclidean conformal field theory that includes
quasi-primary operators corresponding to the principal series. 

The basic axioms of Euclidean quantum field theory were formulated
long-ago by Ostwerwalder and Schraeder. One of the most elementary
axioms needed for a consistent Euclidean theory is that of cluster
decomposition, namely
\[
\lim_{r\to\infty}\left\langle \phi(r)\phi'(0)\right\rangle =\left\langle \phi(r)|0\right\rangle \left\langle 0|\phi'(0)\right\rangle \,,
\]
so that correlators factorize when groups of insertions are separated
by long distance. This is the Euclidean analog of uniqueness of the
vacuum state in Lorentzian signature. It is straightforward to see
this can never be the case for a CFT that contains operators based
on the principal series. Consider the CFT correlator
\[
\left\langle \left(L_{1}\bar{L}_{1}\right)^{n}\mathcal{O}_{\Delta}(z)\left(L_{1}\bar{L}_{1}\right)^{n}\mathcal{O}_{\Delta}^{\dagger}\right\rangle \propto\frac{1}{|z|^{2\Delta-4n}}\,.
\]
This grows with distance for $n>0$, violating cluster decomposition.
If instead one had a typical CFT, and $\mathcal{O}$ was a primary
operator, one would have the identity $L_{n}\mathcal{O}=0$ for $n>1$,
avoiding this problem.

We interpret the results of this paper as a proof by contradiction
that nontrivial CFTs based on the principal series cannot exist. Nevertheless,
this result has important implications for theories in the bulk. In
analogy with AdS/CFT, we can interpret the operator $\left(L_{1}\bar{L}_{1}\right)\mathcal{O}_{\Delta}(z)$
as dual to a composite of a bulk graviton and a scalar matter field.
This violation of cluster decomposition on the boundary arises because
the bulk theory has no positive energy theorem \citep{Abbott:1981ff}.
The Killing vector associated with $L_{0}+\bar{L}_{0}$ is not globally
timelike. There are therefore many bulk excitations satisfying $L_{0}=\bar{L}_{0}=0$
at the boundary, which will appear as intermediate states when one
tries to factorize a CFT correlator.

We conclude then that the Euclidean CFT associated with a free massive
scalar in de Sitter violates the basic axioms of Euclidean quantum
field theory. We take this as a sign that the holographic dual is
incomplete as a CFT, and we hope to return to a more constructive
approach to building the correct holographic dual in future work.

\paragraph*{Acknowledgements}

This research was supported in part by DOE grant DE-SC0010010 and
an FQXi grant. D.L. thanks the Amherst Center for Fundamental Interactions
for hospitality and Jennie Traschen and David Kastor for discussions.

\bibliographystyle{apsrev}
\bibliography{desittir}

\end{document}